\documentclass[a4paper, 10pt]{article}
\usepackage{geometry}
\newcounter{somecounter}
\usepackage{booktabs, array}

\usepackage{setspace}
\usepackage{float}
\usepackage{standalone}
\usepackage{amsmath,amsfonts,amssymb}
\usepackage{amsthm}

\usepackage{color}
\usepackage{xcolor}
\usepackage[english]{babel}

\usepackage{caption}
\usepackage{subcaption}
\usepackage{bm}
\usepackage{pdflscape}
\usepackage{authblk}
\usepackage{graphicx}
\usepackage{hyperref}
\usepackage[normalem]{ulem}
\usepackage{multirow}
\usepackage{multicol}
\usepackage[flushleft]{threeparttable}
\usepackage{enumitem}
\usepackage{url}
\usepackage[utf8]{inputenc}
\usepackage{dsfont}

\usepackage{lipsum}
\makeatletter

\usepackage{algorithm}
\usepackage{algpseudocode}
\usepackage{etoolbox}

\makeatletter
\newenvironment{breakablealgorithm}
  {%
   \begin{center}
     \refstepcounter{algorithm}%
     \hrule height.8pt depth0pt \kern2pt%
     \renewcommand{\caption}[2][\relax]{%
       {\raggedright\textbf{\ALG@name~\thealgorithm} ##2\par}%
       \ifx\relax##1\relax
         \addcontentsline{loa}{algorithm}{\protect\numberline{\thealgorithm}##2}%
       \else
         \addcontentsline{loa}{algorithm}{\protect\numberline{\thealgorithm}##1}%
       \fi
       \kern2pt\hrule\kern2pt
     }%
  }{%
     \kern2pt\hrule\relax%
   \end{center}
  }
\makeatother

\makeatother
\usepackage{natbib}

\usepackage{mathtools}

\newcommand{\interior}[1]{%
 {\kern0pt#1}^{\mathrm{o}}%
}
\usepackage [english]{babel}
\usepackage [autostyle, english = american]{csquotes}
\MakeOuterQuote{"}

\usepackage{algorithm,algpseudocode}
\usepackage{caption}

\usepackage[makeroom]{cancel}

\makeatletter
\newcommand*\bigcdot{\mathpalette\bigcdot@{.5}}
\newcommand*\bigcdot@[2]{\mathbin{\vcenter{\hbox{\scalebox{#2}{$\m@th#1\bullet$}}}}}
\makeatother

\makeatother
\usepackage{natbib}

\usepackage{mathtools}

\usepackage{algorithm,algpseudocode}
\usepackage{caption}
\makeatother

\title{\textbf{A Bayesian Distance-Weighted Ising Model for Task-Related Regional Analysis of Neuroimaging Data}}

\author[1]{Heesang Lee}
\author[1]{Jongrak Park}
\author[5]{Seoyoung Choi}
\author[3]{Bokgyeong Kang}
\author[4,6]{Nial Friel}
\author[5]{Sanghoon Han}
\author[1,2]{Jaewoo Park}
\affil[1]{Department of Statistics and Data Science, Yonsei University}
\affil[2]{Department of Applied Statistics, Yonsei University}
\affil[3]{Department of Statistics, Dongguk University}
\affil[4]{School of Mathematics and Statistics, University College Dublin}
\affil[6]{Rinn Artificial Intelligence}
\affil[5]{Department of Psychology, Yonsei University}

\begin{document}

\maketitle
 
\begin{abstract}
Neuroimaging analysis commonly involves identifying task-relevant brain regions and examining functional connectivity. These objectives are typically addressed separately, without a model-based framework. We propose a Bayesian distance-weighted Ising model that jointly quantifies regional importance for task-label prediction and functional connectivity between brain regions. 
The Ising model favors agreement between interacting labels. We extend it with region-specific interaction parameters and weights based on distances between functional signals. The parameters quantify regional importance for distinguishing tasks, while the weights measure functional connectivity through signal similarity between regions. Inference for this model is challenging because the signals may be observed on unequal time grids and the likelihood function contains an intractable normalizing constant. To address this, we use a basis expansion and an auxiliary variable Markov chain Monte Carlo algorithm. We apply the method to functional near-infrared spectroscopy data from younger and older adults, comparing resting-state and task-related activity. The model reveals age-dependent patterns of regional importance and connectivity. Among younger adults, the left orbitofrontal cortex contributes more strongly to task discrimination than the dorsolateral prefrontal and frontopolar cortices, whereas regional differences are less evident among older adults. Older adults also exhibit greater bilateral connectivity between corresponding regions, consistent with reduced hemispheric lateralization.
\end{abstract}

\noindent%
{\it Keywords: auxiliary variable Markov chain Monte Carlo; functional connectivity; functional near-infrared spectroscopy; Ising model; regional importance}

\section{Introduction}

Neuroimaging techniques such as functional magnetic resonance imaging (fMRI) and functional near-infrared spectroscopy (fNIRS) provide valuable tools for investigating brain activity. By measuring task-related changes in brain signals across multiple cortical locations, researchers can characterize how neural activity varies across experimental conditions and brain regions. Consequently, the statistical analysis of neuroimaging data has become an important component of studies aimed at understanding brain function and its relationship with cognitive processes.

Various methods have been developed for neuroimaging data, with several major analytical objectives, including classification and prediction, where brain signals are used to distinguish between experimental conditions or subject groups. For example, multivoxel pattern analysis (MVPA) treats the decoding of experimental conditions from fMRI signals as a supervised classification problem and has become a widely used approach for analyzing distributed patterns of neural activity \citep{haxby2001distributed, norman2006beyond}. Related classification approaches have also been extensively studied for fNIRS data, particularly in brain--computer interface, clinical diagnosis, and cognitive-task applications \citep{naseer2015fnirs}. 
Another major objective is to identify brain regions that contribute most strongly to distinguishing different cognitive states or tasks \citep{haufe2014interpretation}. 
Such analyses can be summarized through spatial importance maps, which highlight regions associated with task discrimination. Another major objective is connectivity analysis \citep{bullmore2009complex, zhang2010functional}, which investigates relationships among different brain regions and reveals coordinated patterns of neural activity. 

Several statistical methods have been developed to identify brain regions that contribute to predicting clinical or behavioral outcomes from neuroimaging data. \citet{filippone2012probabilistic} use a Bayesian multinomial logistic regression model to distinguish among neurological disorders and assess the relative contributions of imaging modalities and brain regions to classification. \citet{goldsmith2014smooth} develop a Bayesian scalar-on-image regression model that accounts for spatial dependence among neighboring voxels to identify brain regions associated with scalar outcomes. \citet{li2015spatial} adopt a related Bayesian framework and use a Dirichlet process prior to group voxels with similar regression effects. Although these approaches assess the contributions of brain regions to predicting clinical or behavioral outcomes, they do not explicitly characterize functional connectivity between regions. In contrast, \citet{relion2019network} propose a penalized classification approach that uses functional connectivity networks to distinguish between subject groups and select brain regions and connections that contribute to classification. However, this approach relies on precomputed connectivity measures and does not explicitly quantify uncertainty in regional importance.

Motivated by this limitation, we propose a Bayesian distance-weighted Ising (BDWI) model, which is a unified probabilistic model for analyzing functional neuroimaging data. Our primary aim is to characterize regional importance and inter-regional signal similarity. Rather than using precomputed connectivity measures, our model uses distances between functional observations to quantify dependence among condition labels. Region-of-interest (ROI)-specific interaction parameters allow observations from different brain regions to contribute differently to condition-label prediction. Bayesian inference provides estimates of these interaction strengths and model-based summaries of inter-regional similarity, together with uncertainty quantification for comparisons across ROIs.

Our approach builds on probabilistic nearest-neighbor methods for classification. \citet{holmes2002probabilistic} introduced a probabilistic formulation of the $k$-nearest-neighbor ($k$-NN) classifier, but the conditional distributions of the class labels generally do not correspond to a valid joint distribution.
\citet{cucala2009bayesian} addressed this limitation by defining a valid joint distribution for the class labels. Their model retains a neighborhood structure defined by the $k$ nearest observations, with observations within the neighborhood contributing equally regardless of their distances. For two classes, this joint distribution takes the form of an Ising model. However, both models require the number of neighbors $k$ to be specified, and their performance depends on the choice of $k$. \citet{friel2011classification} introduced the distance nearest-neighbor (d-NN) model, which assigns distance-based weights to all other observations, avoiding the need to specify $k$. However, directly applying the d-NN model to functional neuroimaging data presents two important challenges. First, the original d-NN model assumes a common interaction parameter across all observations and therefore does not account for regional heterogeneity. In functional neuroimaging data, the contribution of functional observations to task-label prediction may vary across brain regions. Second, neuroimaging signals are high-dimensional, with each functional observation consisting of a large number of time points. Moreover, the number and timing of measurements may vary across observations or experimental conditions, complicating the direct computation of distances between the observed time series.

To address these challenges, we account for regional heterogeneity by introducing ROI-specific interaction parameters. We also employ basis expansion to obtain a common low-dimensional representation of the functional signals, enabling distance-based comparisons between high-dimensional observations measured on different grids. 
The likelihood for our proposed model contains an intractable, parameter-dependent normalizing constant. Thus, standard Markov chain Monte Carlo (MCMC) algorithms cannot be applied directly for Bayesian inference.
We address this computational difficulty using an auxiliary variable MCMC \citep{murray2006, ciamo-friel2011} approach that avoids direct evaluation of the normalizing constant.

The remainder of this paper is organized as follows. In Section~\ref{sec:fnirs}, we describe the fNIRS data from younger and older adults and present exploratory analyses. In Section~\ref{sec:method}, we introduce our BDWI model and describe an auxiliary variable MCMC algorithm for posterior inference. In Section~\ref{sec:application}, we apply our proposed method to the fNIRS data and examine regional importance and distance-weighted functional connectivity. In Section~\ref{sec:discussion}, we conclude the paper with a discussion and summary.

\section{Functional Near-Infrared Spectroscopy Data}
\label{sec:fnirs}

In this section, we describe the fNIRS data and use exploratory analyses to motivate our model by examining how task-related patterns vary across brain regions and age cohorts.

\subsection{Data Description}
\label{sec:data}

The fNIRS data consist of time-series measurements of changes in oxygenated and deoxygenated hemoglobin concentrations and provide information about cortical hemodynamic activity. The fNIRS recordings were collected in the Department of Psychology at Yonsei University from two separately recruited cohorts of younger and older adults. The younger cohort included participants aged 19--33 years, whereas the older cohort consisted of participants aged 65 years or older. Younger adults were recruited from four-year universities in South Korea and were either current undergraduate students or recent graduates. Older adults were recruited from a
senior program in Seongnam-si, Gyeonggi-do. For older adults, cognitive impairment was screened using the Korean Mini-Mental State Examination (K-MMSE). In both cohorts, participants were excluded if they had a current or previous Axis~I psychiatric disorder, were taking medication that could alter neural activity, or had uncorrected visual impairment. In this manuscript, we analyze the two cohorts separately. Cortical hemodynamic activity was recorded in the prefrontal cortex using a continuous-wave fNIRS system (NIRSIT; OBELAB Inc., Seoul, South Korea), which consists of 48 measurement channels sampled at 8.138 Hz. Each channel records changes in oxygenated (HbO) and deoxygenated (HbR) hemoglobin concentrations, and we use the HbO signal for the subsequent analysis. The recordings were preprocessed following standard fNIRS procedures \citep{takizawa2014neuroimaging, pfeifer2018signal, Yucel2021}. 

\begin{figure}[htbp]
\centering
    {\includegraphics[width=0.7\linewidth]{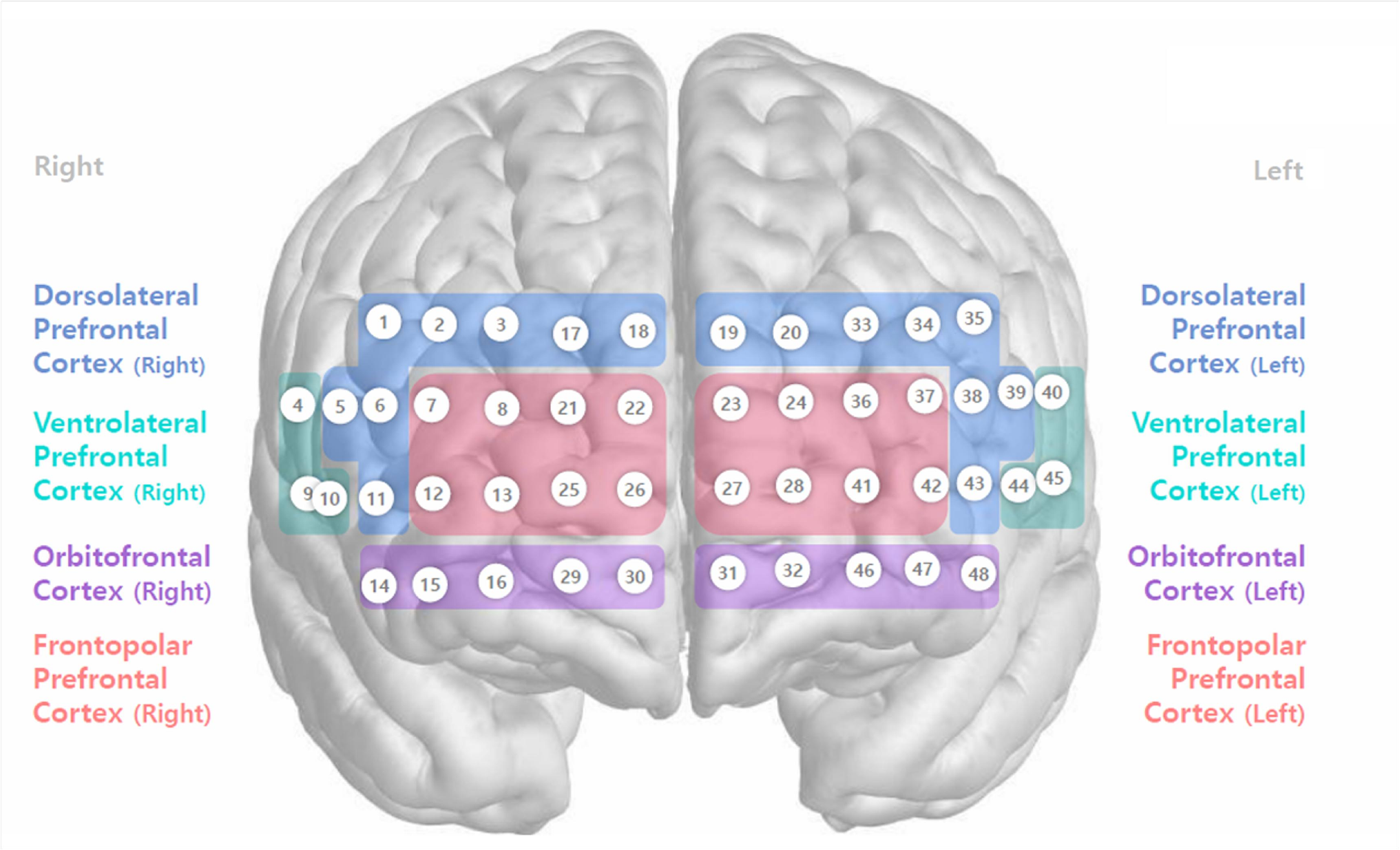}}
    {\caption{Anatomical partition of the prefrontal cortex into eight ROIs. Each of the 48 fNIRS measurement channels is assigned to one ROI according to its cortical location.}
    \label{fig:roi8_regions}}
\end{figure}

Each participant completed three experimental conditions: (1) a resting-state condition, during which the participant sat quietly with their eyes closed; (2) the Iowa gambling task (IGT), a repeated-choice task designed to assess decision-making under uncertainty \citep{Bechara1994}; and (3) the Stroop task, a color--word interference task that assesses cognitive control and selective attention \citep{Stroop1935}. Detailed information on participant characteristics, experimental design and procedures, task protocols, and fNIRS data acquisition is provided in Section~1 of the Supplementary Material. Our interest lies in how cortical activity changes when a participant engages in a cognitive task relative to rest. 
We therefore compare the IGT and Stroop conditions separately with the resting state to characterize task-related differences in cortical activity.
In our study, we assign each of the 48 measurement channels to one of eight anatomically defined ROIs according to its cortical location, using Brodmann area (BA) mapping \citep{brodmann1909vergleichende}.
Thus, this grouping does not select or aggregate the 48 channel-level signals into eight observations; rather, it partitions the 48 channels into eight broader cortical regions while retaining each channel as an individual functional observation.
The eight ROIs comprise the dorsolateral prefrontal cortex (DLPFC), ventrolateral prefrontal cortex (VLPFC), frontopolar prefrontal cortex (FPC), and orbitofrontal cortex (OFC) in both hemispheres. Specifically, the left-hemisphere ROIs are denoted by L-DLPFC (ROI 1L), L-VLPFC (ROI 2L), L-FPC (ROI 3L), and L-OFC (ROI 4L), while the corresponding right-hemisphere ROIs are denoted by R-DLPFC (ROI 1R), R-VLPFC (ROI 2R), R-FPC (ROI 3R), and R-OFC (ROI 4R) (Figure~\ref{fig:roi8_regions}). These regions provide an anatomically interpretable partition of the prefrontal cortex and are used to characterize regional variation in the relationship between the fNIRS signals and experimental conditions. 

\begin{figure}[htbp]
\centering
    {\includegraphics[width=0.95\linewidth]{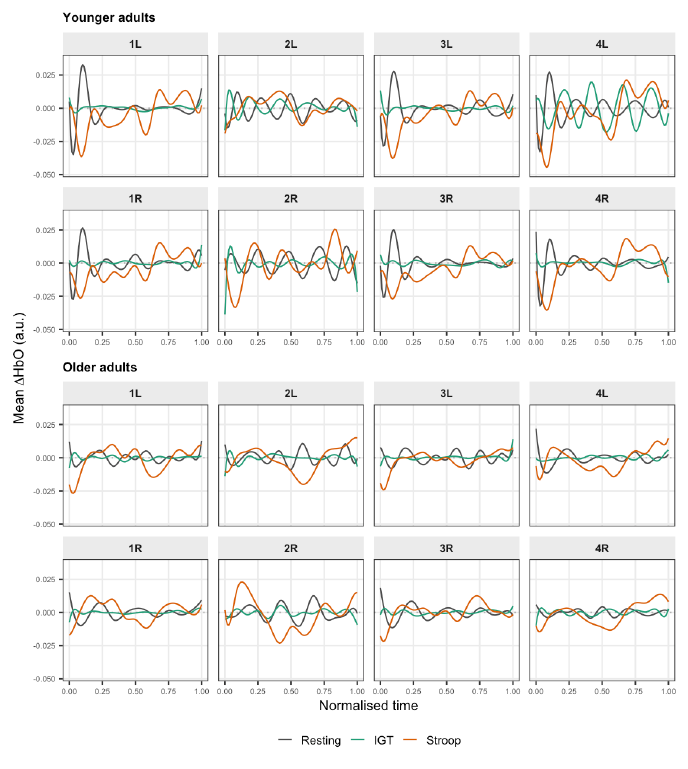}}
    {\caption{Mean HbO curves for each condition within each ROI, shown separately for younger and older adults. The resulting curves are averaged over participants and channels within each ROI, condition, and age cohort.
    }
    \label{fig:eda_roi8_task_young&old}}
\end{figure}

\begin{figure}[htbp]
\centering
    {\includegraphics[width=0.8\linewidth]{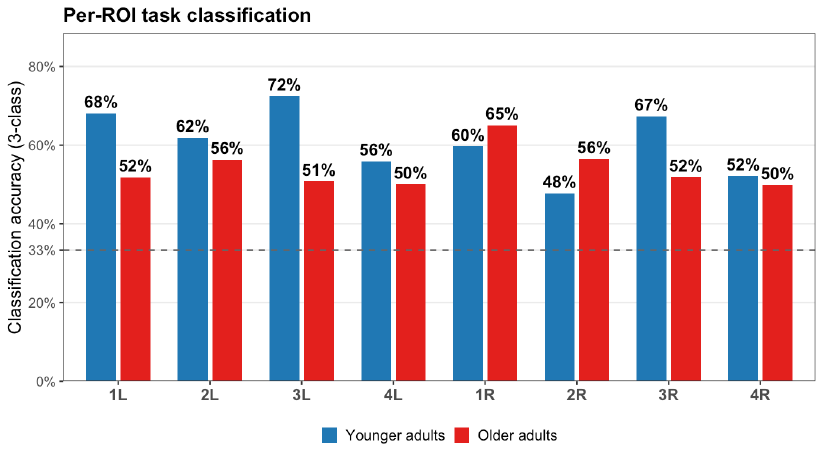}}
    {\caption{Three-class classification accuracy within each ROI and age cohort using a $ k$-nearest-neighbor classifier ($k$ = 25). Accuracy is evaluated by 5-fold cross-validation, with all observations from a given participant assigned to the same fold. Blue and red bars represent younger and older adults, respectively. The horizontal dashed line indicates the expected accuracy under uniform random assignment to the three experimental conditions ($1/3$).}
    \label{fig:eda_roi8_knn}}
\end{figure}

\subsection{Exploratory Data Analysis}
\label{sec:data_eda_roi8}

We conduct an exploratory data analysis to motivate our model by examining regional and cohort-specific variation in task-related HbO signals. We first compare the task-specific mean HbO curves within each ROI. For each curve, we normalize the time domain to $[0,1]$ and linearly interpolate the HbO time series at 100 equally spaced time points. This transformation places recordings of different durations—approximately 300, 856, and 63 seconds for the resting, IGT, and Stroop conditions, respectively—on a common time grid. For each ROI–condition combination, we compute the mean curve by averaging the corresponding channel-by-participant observations at each normalized time point, as illustrated in Figure~\ref{fig:eda_roi8_task_young&old}.
We observe that the temporal patterns differ across conditions and ROIs, with the Stroop task showing particularly distinct trajectories compared with the resting and IGT conditions. Corresponding left--right ROIs generally exhibit similar patterns, although their magnitudes vary across ROIs and cohorts. 
Across most ROIs, the mean HbO curves show larger temporal fluctuations in the younger cohort than in the older cohort. In the older cohort, differences between Stroop and the other conditions remain apparent despite the smaller overall amplitude range. These findings motivate a model that accommodates ROI-specific heterogeneity.
 

Furthermore, we examine the extent to which individual ROIs are informative for distinguishing among the three experimental conditions. For each ROI and cohort separately, we train a $ k$-nearest-neighbor classifier ($k=25$) using Euclidean distances between the curves, with each observation corresponding to a participant–channel–condition combination. Performance is evaluated using 5-fold cross-validation. Figure~\ref{fig:eda_roi8_knn} shows the held-out classification accuracies, with $1/3$ 
denoting the expected accuracy under uniform random assignment to the three experimental conditions.
All ROIs achieve accuracies above this baseline, although accuracy varies across ROIs and cohorts. In the younger cohort, ROI~3L achieves the highest accuracy ($72\%$), followed by ROI~1L ($68\%$) and ROI~3R ($67\%$), whereas ROI~1R performs best in the older cohort ($65\%$). 
Classification accuracy is higher in older adults for ROIs 1R and 2R and in younger adults for the remaining ROIs.
Together, these exploratory analyses suggest that the extent to which HbO signals distinguish between experimental conditions varies across ROIs and age cohorts.


\section{Method}
\label{sec:method}

This section introduces our proposed model that accommodates heterogeneity across ROIs to assess regional importance and characterize functional connectivity between brain regions. We then describe how to perform Bayesian inference, interpret of the model parameters, and assess model fit. 

\subsection{A Bayesian Distance-Weighted Ising Model}

Let $X_{ijk}(t)$ denote the observed HbO signal for participant $i$, channel $j$, and condition label $k$, where $i=1,\ldots,N$, $j=1,\ldots,48$, and $k\in\{0,1\}$. Since our primary interest is in distinguishing resting state activity from task-related activity, we consider a binary classification setting. Here, $k=0$ denotes the resting state, and $k=1$ denotes the corresponding task (e.g., Stroop or IGT). For each participant and experimental condition, the HbO signal is observed at time points $t\in\mathcal{T}$. Note that the experimental conditions have different durations, and the usable observation periods may vary across recordings after preprocessing; the functional signals are not necessarily observed on a common time grid. Moreover, the fNIRS signals are recorded at densely sampled time points, resulting in a high-dimensional representation. To obtain a common low-dimensional representation across unequal observation grids, we use the following basis expansion:
\begin{equation}
X_{ijk}(t)
\approx
\sum_{m=1}^{M} x_{ijkm}\phi_m(t),
\qquad t\in\mathcal{T},
\label{eq}
\end{equation}
where ${\phi_m(t)}$ is the $m$th basis function and $x_{ijkm}$ its coefficient. In our study, the number of basis functions is selected by minimizing the generalized cross-validation criterion, resulting in $M=19$ B-spline basis functions. 

For subsequent analysis, let $\ell=1,\ldots,n$ index all participant-channel-condition combinations $(i,j,k)$. For the $\ell$th functional observation, let
$\mathbf{x}_{\ell}=(x_{\ell 1},\ldots,x_{\ell M})^\top\in\mathbb{R}^M$
denote the vector of basis coefficients representing its functional signal. Collect these vectors in $\mathbf x=(\mathbf{x}_1,\ldots,\mathbf{x}_n)$. Let $y_\ell$ denote the condition label associated with the $\ell$th observation and $\mathbf y=(y_1,\ldots,y_n) \in \{0,1\}^{n}$. In our study, for each experiment containing 40 participants, 48 channels, and two experimental conditions, the total number of functional observations is $n=40\times48\times2=3,840$.

To analyze the functional neuroimaging data, we propose a Bayesian distance-weighted Ising (BDWI) model that characterizes task labels using connectivities between functional observations. Our approach builds on the d-NN model \citep{friel2011classification}, an Ising-type classifier that weights pairwise interactions by feature-space distance. We extend d-NN by introducing ROI-specific interaction parameters to account for heterogeneity in task-related patterns across brain regions. The joint distribution of the labels is given by
\begin{equation} 
p(\mathbf y | \mathbf x, \boldsymbol\beta,\sigma) = \frac{ \exp\left\{ S(\mathbf y;\mathbf x, \boldsymbol\beta,\sigma) \right\} }{ Z(\boldsymbol\beta,\sigma) }, \label{eq:dnn_mrf_joint_theta_final} 
\end{equation}
where 
\[
S(\mathbf y;\mathbf x, \boldsymbol\beta,\sigma) = \sum_{\ell=1}^{n} \sum_{\substack{\ell'=1 \\ \ell'\neq \ell}}^{n} \beta_{r_\ell} w_{\ell \ell'}(\sigma) \mathbb I(y_\ell=y_{\ell'}), 
\] 
and $Z(\boldsymbol\beta, \sigma)$ is a normalizing constant. Here, $\boldsymbol\beta=(\beta_1,\ldots,\beta_{8})^\top$, with $\beta_j>0$, denotes the ROI-specific interaction parameters, where $\beta_j$ corresponds to the $j$th ROI, and $r_\ell\in\{1,\ldots,8\}$ indicates the ROI to which observation $\ell$ belongs.
The parameter $\beta_{r_\ell}$ controls the strength of distance-based interactions associated with ROI $r_\ell$. Larger values of $\beta_{r_\ell}$ increase the influence of training observations from ROI $r_\ell$ on condition-label prediction. For a new observation, training observations from this ROI with similar functional signals support a resting-state prediction if they predominantly belong to the resting condition. If they predominantly belong to the task condition, they support prediction of that condition.
The ROI-specific parameters thus capture regional heterogeneity in task classification. For each pair of observations $(\ell,\ell')$, we define the weight as
\begin{equation}
w_{\ell\ell'}(\sigma)
\propto
\exp\left\{
-\frac{d(\mathbf{x}_\ell,\mathbf{x}_{\ell'})}{2\sigma^2}
\right\},
\label{eq:dnn_weights_roi_theta}
\end{equation}
where $d(\cdot,\cdot)$ denotes a Euclidean distance between the coefficient vectors and $\sigma>0$ is a bandwidth parameter. The weights assign larger values to nearby observations and smaller values to distant observations. The bandwidth parameter $\sigma$ controls the rate of this decay: smaller values place greater emphasis on nearby observations, whereas larger values allow more distant observations to contribute to the interaction. 

To model ROI-specific interaction, we decompose the log-interaction parameters as 
\[
\log\beta_{r_\ell} = \alpha + u_{r_\ell},  
\]
where $\alpha$ is the global interaction level and $u_{r_\ell}$ is the deviation specific to ROI $r_\ell$. For identifiability, we impose the sum-to-zero constraint $\sum_{r_\ell=1}^{8} u_{r_\ell} = 0$. Consequently,
\[
\alpha=\overline{\log\beta}
=\frac{1}{8}\sum_{r=1}^{8}\log\beta_r,
\qquad
u_r=\log\beta_r-\overline{\log\beta}.
\]
We assign a Gaussian prior to the global component, $\alpha \sim \mathcal{N}(-1,0.2),$ and a centred Gaussian prior to the ROI-specific deviations $\mathbf u=(u_1,\ldots,u_8)'$, where $u_{r_\ell}$ represents the deviation of ROI $r_\ell$ from the global component,
\[
p(\mathbf u | \tau^2)
\propto
\exp\left\{
-\frac{1}{2\tau^2}\sum_{r=1}^{8}u_r^2
\right\},
\qquad
\sum_{r=1}^{8}u_r=0.
\]
Equivalently, this induces the following prior on $\log\boldsymbol\beta$:
\[
p(\log\boldsymbol\beta|\tau^2)
\propto
\exp\left\{
-\frac{(\overline{\log\beta}+1)^2}{0.4}
-\frac{1}{2\tau^2}
\sum_{r_\ell=1}^{8}
(\log\beta_{r_\ell}-\overline{\log\beta})^2
\right\}.
\]
The variance parameter $\tau^2$ controls the degree of heterogeneity among the ROI-specific interaction parameters. Smaller values of $\tau^2$ encourage the latent parameters to remain close to their common level, resulting in more similar interaction strengths across ROIs, whereas larger values allow greater ROI-specific variation. 
We assign the following hyperpriors to the variance parameter governing ROI heterogeneity and to the bandwidth parameter controlling the neighborhood weights: \[\tau^2 \sim \mathrm{IG}(2,1), \qquad \log \sigma \sim \mathcal{N}(0.5,0.2). \]

\subsection{Bayesian Inference}
\label{sec:bayesian}

In \eqref{eq:dnn_mrf_joint_theta_final}, evaluation of $Z(\boldsymbol\beta,\sigma)=\sum_{\mathbf y\in\{0,1\}^{n}}
\exp\left\{S(\mathbf y;\mathbf x,\boldsymbol\beta,\sigma)\right\}$ requires summation over all $2^n$ possible configurations of the class labels, which is intractable even for moderate values of $n$. Under the hierarchical specification, the resulting posterior distribution is \begin{equation}
p(\log\boldsymbol\beta,\tau^2, \log\sigma | \mathbf x, \mathbf y)
\propto
\frac{\exp\{S(\mathbf y;\mathbf x,\boldsymbol\beta,\sigma)\}}
{Z(\boldsymbol\beta,\sigma)}
p(\log\boldsymbol\beta | \tau^2)
p(\tau^2)
p(\log\sigma),
\label{eqn:target-posterior}
\end{equation}
which is a doubly intractable distribution \citep{murray2006}. Since $Z(\boldsymbol\beta,\sigma)$ is intractable, the standard Metropolis-Hastings acceptance probability cannot be evaluated \citep{haran2026algorithms}.

To address this issue, we employ the approximate exchange algorithm (AEA)
\citep{murray2006, ciamo-friel2011}. Given the current value $\log\boldsymbol\beta$, we first propose
$\log\boldsymbol\beta' \sim q(\log\boldsymbol\beta' \mid \log\boldsymbol\beta),$
and obtain the corresponding interaction parameters as
$\boldsymbol\beta'=\exp(\log\boldsymbol\beta')$.
Conditional on $\boldsymbol\beta'$, an auxiliary Markov chain is run with
stationary distribution
\[
p(\mathbf y|\mathbf x,\boldsymbol\beta',\sigma)
=
\frac{
    \exp\{S(\mathbf y;\mathbf x,\boldsymbol\beta',\sigma)\}
}{
    Z(\boldsymbol\beta',\sigma)
}.
\]
Starting from $\mathbf y^{(0)}$, we run the auxiliary chain for $L$ Gibbs
sweeps and use its final state, $\mathbf y'=\mathbf y^{(L)}$, as an
approximate draw from
$p(\cdot|\mathbf x,\boldsymbol\beta',\sigma)$. Here, one Gibbs sweep
consists of sequentially updating every component of the auxiliary
configuration from its full conditional distribution. The proposed value $\boldsymbol\beta'$ is accepted with probability
\[\min\left\{\frac{\exp{S(\mathbf y;\mathbf x,\boldsymbol\beta',\sigma)}\exp{S(\mathbf y';\mathbf x,\boldsymbol\beta,\sigma)}p(\log\boldsymbol\beta' | \tau^2)q(\log\boldsymbol\beta | \log\boldsymbol\beta')\cancel{Z(\boldsymbol\beta,\sigma)}\cancel{Z(\boldsymbol\beta',\sigma)}}{\exp{S(\mathbf y;\mathbf x,\boldsymbol\beta,\sigma)}\exp{S(\mathbf y';\mathbf x,\boldsymbol\beta',\sigma)}p(\log\boldsymbol\beta | \tau^2)q(\log\boldsymbol\beta' | \log\boldsymbol\beta)\cancel{Z(\boldsymbol\beta',\sigma)}\cancel{Z(\boldsymbol\beta,\sigma)}}, 1\right\},\]
where the intractable normalizing constants cancel out. If the auxiliary variable can be sampled exactly from the
likelihood model, for example through perfect sampling, the resulting
exchange algorithm targets the posterior distribution in
\eqref{eqn:target-posterior} exactly. With a finite number of Gibbs sweeps, the algorithm is approximate; \citet{everitt2012} provides theoretical justification for its convergence as $L$ increases. The bandwidth parameter $\sigma$ is updated analogously. Since $\tau^2$ does not appear in $Z(\boldsymbol\beta,\sigma)$, it can be directly updated from the conditional distribution. The entire procedure is summarised in Algorithm~\ref{alg:BDWI}.



\begin{breakablealgorithm}
\caption{An MCMC Algorithm for our proposed BDWI Model}
\label{alg:BDWI}
\begin{algorithmic}[1]

\Require
Observed data $\mathcal D=\{(\mathbf{x}_\ell,y_\ell,r_\ell)\}_{\ell=1}^n$, number of MCMC iterations $T$, auxiliary Gibbs sweeps $L$, initial values $\boldsymbol\beta^{(0)}, \sigma^{(0)}$, proposal variances $v_\sigma^2$, and $v_\beta^2.$ 

\vspace{1mm}

\For{$t=1,\ldots,T$}

\Statex \textbf{Step 1: Update for $\sigma$}

\State Propose $\log \sigma'
=\log \sigma^{(t-1)}
+\varepsilon_\sigma,
\qquad
\varepsilon_\sigma \sim \mathcal N(0,v_\sigma^2).$

\State Generate $\mathbf y'$ using $L$ Gibbs sweeps targeting
$p(\cdot | \mathbf x,\boldsymbol\beta^{(t-1)}, \sigma')$.

\State Accept $\sigma^{(t)}=\sigma'$ with probability 
\[
\min\left\{ \frac{\exp\lbrace S(\mathbf y; \mathbf x, \boldsymbol\beta^{(t-1)}, \sigma')\rbrace\exp\lbrace S(\mathbf y';\mathbf x, \boldsymbol\beta^{(t-1)}, \sigma^{(t-1)}) \rbrace p(\log\sigma')}{\exp\lbrace S(\mathbf y;\mathbf x, \boldsymbol\beta^{(t-1)}, \sigma^{(t-1)})\rbrace \exp\lbrace S(\mathbf y';\mathbf x, \boldsymbol\beta^{(t-1)},\sigma')\rbrace p(\log\sigma^{(t-1)})}, 1\right\}
\]
else reject (set $\sigma^{(t)}=\sigma^{(t-1)}$).

\vspace{1mm}

\Statex \textbf{Step 2: Update for $\boldsymbol \beta$}

\For{$r_\ell=1,\ldots,8$}

\State Propose $\log\beta_{r_\ell}'=\log\beta_{r_\ell}^{(t-1)}+
\varepsilon_{r_\ell},\qquad
\varepsilon_{r_\ell}\sim \mathcal N(0,v_\beta^2).$

\State Set $\beta_{r_\ell}'=\exp(\log\beta_{r_\ell}')$.

\State Generate $\mathbf y'$ using $L$ Gibbs sweeps targeting
$p(\cdot | \mathbf x,\boldsymbol\beta',\sigma^{(t)})$.

\State Accept $\beta_{r_\ell}^{(t)}=\beta_{r_\ell}'$ with probability 
\[
\min\left\{ \frac{\exp\lbrace S(\mathbf y; \mathbf x, \boldsymbol\beta', \sigma^{(t)}) \rbrace \exp\lbrace S(\mathbf y';\mathbf x, \boldsymbol\beta^{(t-1)}, \sigma^{(t)}) \rbrace p(\log\boldsymbol\beta' | \tau^{2})}{\exp\lbrace S(\mathbf y;\mathbf x,\boldsymbol\beta^{(t-1)},\sigma^{(t)})\rbrace \exp\lbrace S(\mathbf y';\mathbf x,\boldsymbol\beta',\sigma^{(t)}) \rbrace p( \log\boldsymbol\beta^{(t-1)} | \tau^{2})}, 1\right\}
\]
else reject (set $\beta_{r_\ell}^{(t)}=\beta_{r_\ell}^{(t-1)}$).

\EndFor

\vspace{1mm}

\Statex \textbf{Step 3: Update for $\tau^2$}

\State Sample 
\[
\tau^{2(t)} \sim \mathrm{IG} \left(a_\tau+\frac{7}{2},\, b_\tau+\frac12\sum_{r_\ell=1}^{8}(u_{r_\ell}^{(t)})^2\right),
\]
where $u_{r_\ell}^{(t)}=\log\beta_{r_\ell}^{(t)}-\overline{\log\beta^{(t)}}$ and 
$\overline{\log\beta^{(t)}} = \frac{1}{8}\sum_{r_\ell=1}^{8}\log\beta^{(t)}_{r_\ell}$.

\vspace{1mm}

\EndFor
\end{algorithmic}
\end{breakablealgorithm}

\pagebreak[2]
\subsection{Neuroimaging Interpretation}

Our BDWI model supports three complementary analyses of neuroimaging data. First, the model enables task classification based on fNIRS data. This demonstrates that our model can capture task-specific patterns in neuroimaging signals and use them to distinguish among different experimental conditions.

Second, by estimating ROI-specific interaction strengths $\beta_{r_\ell}$, the model provides an ROI-level measure of each ROI's contribution to task-label prediction. 
A larger $\beta_{r_\ell}$ strengthens distance-weighted interactions between observations from ROI $r_\ell$ and other observations, whether they belong to the same or different ROIs. These interactions favor agreement in condition labels between observations with similar functional signals.
An importance map of $\beta_{r_\ell}$ identifies ROIs that are most informative for task discrimination.

Lastly, the weights $w_{\ell\ell'}$ in \eqref{eq:dnn_weights_roi_theta}, which quantify connectivity between neuroimaging time-series signals, can be used to characterize connectivity between ROIs. We define the distance-weighted connectivity between ROIs $j$ and $k$ as
\[
\overline W_{jk}(\sigma)
=
\frac{1}{|A_j|\,|A_k|}
\sum_{\ell\in A_j}
\sum_{\ell'\in A_k}
w_{\ell \ell'}(\sigma),
\qquad
A_j=\{\ell:\, r_\ell=j\}.
\]
For interpretability, we standardize this as:
\begin{equation}
\widetilde C_{jk}(\sigma)
=
\frac{
\overline W_{jk}(\sigma)-\mu_W(\sigma)
}{
\sqrt{v_W(\sigma)}
},
\qquad j<k,
\label{eq:connectivity}
\end{equation}
where $\mu_W(\sigma)=\frac{1}{28}
\sum_{a<b} \overline W_{ab}(\sigma)$ denotes the mean pairwise connectivity, and $v_W(\sigma)=
\frac{1}{27}\sum_{a<b}
\left\{\overline W_{ab}(\sigma)-\mu_W(\sigma)
\right\}^{2}$ denotes the corresponding sample variance.
An estimate of $\widetilde C_{jk}$ can be obtained from the posterior samples of $\sigma$. A positive value indicates that the ROI pair exhibits greater-than-average connectivity, whereas a negative value indicates lower-than-average connectivity. Its magnitude represents the deviation
from the overall mean in units of the standard deviation.

\subsection{Model Assessment}

To assess model fit, we conduct posterior predictive checks (PPCs) based on ROI-specific contributions to the model statistic $S(\mathbf y;\mathbf x,\boldsymbol\beta,\sigma)$. For each retained posterior draw
$(\boldsymbol\beta^{(t)},\sigma^{(t)})$, we generate a posterior predictive (pp) sample $\mathbf y^{\mathrm{pp},(t)}$ from
$p(\cdot | \mathbf x,\boldsymbol\beta^{(t)},\sigma^{(t)})$ using $L=10$ Gibbs sweeps. We express the model statistic as a sum of ROI-specific contributions:
\[S(\mathbf y;\mathbf x,\boldsymbol\beta,\sigma)=\sum_{r=1}^8S_r(\mathbf y;\mathbf x,\boldsymbol\beta,\sigma),\]
where 
\[S_r(\mathbf y;\mathbf x,\boldsymbol\beta,\sigma)=\sum_{\ell:r_\ell=r} \sum_{\substack{\ell'=1 \\ \ell'\neq \ell}}^{n} \beta_{r_\ell} w_{\ell \ell'}(\sigma) \mathbb I(y_\ell=y_{\ell'}).\]
For $r=1,\ldots,8$, we compute the posterior predictive $p$-value as 
\[
\widehat{p}_r=\frac{1}{B}\sum_{b=1}^{B}
\mathbb I
\left[S_r(\mathbf y^{\mathrm{pp},(b)};\mathbf x,
\boldsymbol\beta^{(b)},\sigma^{(b)})\ge
S_r(\mathbf y;\mathbf x,
\boldsymbol\beta^{(b)},\sigma^{(b)})\right],
\]
where $B$ denotes the number of retained posterior draws. Values of $\widehat{p}_r$ near $0.5$ indicate that the observed ROI-specific contribution is consistent with those obtained from posterior predictive samples, whereas values near 0 or 1 indicate
potential lack of fit. To summarize the ROI-specific posterior predictive assessments in a single measure, we define
\begin{equation}
D_{\mathrm{PPC}}=\sqrt{\sum_{r=1}^8(\widehat{p}_r-0.5)^2}.  
\label{eq:dev}
\end{equation}
A smaller value of $D_{\mathrm{PPC}}$ indicates a better model fit.

\section{Application}
\label{sec:application}

In this section, we apply our proposed BDWI model to the fNIRS data. We consider two binary classification problems, comparing the IGT with the resting state and comparing the Stroop task with the resting state. Each comparison is conducted separately for the younger and older cohorts, resulting in a total of four datasets. Inference is performed using the AEA described in  Section~\ref{sec:bayesian}. For each dataset, we run the algorithm for 20,000  iterations, discard the first 10,000 iterations as burn-in, and retain every 10th draw thereafter. Following the heuristic of \citet{park2018bayesian}, we set \(L=10\), beyond which the posterior distributions change little. The sensitivity analysis in Section~2 of the Supplementary Material shows accurate parameter recovery for both $L=10$ and $L=20$.

\subsection{Model Comparison}

We fit our model and several baseline models to the data and compare their fit using $D_{\mathrm{PPC}}$ in \eqref{eq:dev}. The d-NN model \citep{friel2011classification} uses a homogeneous interaction parameter $\beta$ across all ROIs while retaining the distance-based weights:
\[S_{\mathrm{d-NN}}(\mathbf y;\mathbf x,\boldsymbol\beta,\sigma)= \beta \sum_{\ell=1}^n \sum_{\substack{\ell'=1 \\ \ell'\neq \ell}}^{n}  w_{\ell \ell'}(\sigma) \mathbb I(y_\ell=y_{\ell'}).\]
The non-spatial model does not use any distance or neighborhood information and instead assigns equal weight to all pairs of observations:
\[S_{\mathrm{NS}}(\mathbf y;\mathbf x,\boldsymbol\beta,\sigma)= \beta \sum_{\ell=1}^n \sum_{\substack{\ell'=1 \\ \ell'\neq \ell}}^{n}  \mathbb I(y_\ell=y_{\ell'}).\]
In addition, as an extension of our proposed model, we consider ROI-specific bandwidth parameters. Specifically, the common bandwidth parameter $\sigma$ in the main model is replaced by $\boldsymbol{\sigma}=(\sigma_1,\ldots,\sigma_8)'$, where $\sigma_{r_\ell}>0$ denotes the bandwidth parameter associated with ROI $r_\ell$:
\[S_{\boldsymbol{\sigma}}(\mathbf y;\mathbf x,\boldsymbol\beta,\boldsymbol\sigma)
=
\sum_{\ell=1}^{n}
\sum_{\substack{\ell'=1\\ \ell'\neq\ell}}^{n}
\beta_{r_\ell}
w_{\ell\ell'}(\sigma_{r_\ell})
\mathbb I(y_\ell=y_{\ell'}).\]
We refer to the extended version as BDWI ($\boldsymbol{\beta}, \boldsymbol{\sigma}$) and the original model as BDWI ($\boldsymbol{\beta}$).
\begin{table*}[ht]
\caption{$D_{\mathrm{PPC}}$ values for our BDWI, d-NN, and non-spatial models, reported separately for the IGT versus resting and Stroop versus resting comparisons in the younger and older cohorts.}
\label{tab:classification}
\tabcolsep=0pt
\begin{tabular*}{\textwidth}{@{\extracolsep{\fill}}lcccc@{\extracolsep{\fill}}}
\toprule%
& \multicolumn{2}{@{}c@{}}{Younger} 
& \multicolumn{2}{@{}c@{}}{Older} \\
\cline{2-3}\cline{4-5}%
Model 
& IGT vs. Resting 
& Stroop vs. Resting 
& IGT vs. Resting 
& Stroop vs. Resting \\
\midrule
BDWI ($\boldsymbol{\beta}$) 
& 0.112 
& 0.122 
& 0.196 
& 0.126 \\
BDWI ($\boldsymbol{\beta}, \boldsymbol{\sigma}$) 
& 0.103 
& 0.101 
& 0.217 
& 0.083 \\
d-NN 
& 0.348 
& 0.198 
& 0.747 
& 0.137 \\
Non-spatial 
& 0.648 
& 0.581 
& 0.631 
& 0.631 \\
\bottomrule
\end{tabular*}
\end{table*}

Table~\ref{tab:classification} summarizes $D_{\mathrm{PPC}}$ for each model. Our BDWI model consistently achieves lower $D_{\mathrm{PPC}}$ values than d-NN, indicating that ROI-specific interaction parameters better capture the regional characteristics of the observed data. The extended BDWI ($\boldsymbol{\beta}, \boldsymbol{\sigma}$) model shows no substantial improvement in model fit over BDWI ($\boldsymbol{\beta}$). We therefore retain the more parsimonious BDWI ($\boldsymbol{\beta}$) model for the main analysis.


\subsection{Importance Analysis}
We investigate the relative importance of individual brain regions in distinguishing each cognitive task from the resting state. In our proposed BDWI model, the ROI-specific interaction parameter $\beta_{r_\ell}$ controls the strength of local label agreement associated with ROI ${r_\ell}$. Accordingly, inference on this parameter measures the relative contribution of each ROI to task discrimination. Larger values of $\beta_{r_\ell}$ indicate stronger task-discriminative local dependence associated with the corresponding ROI. For each ROI, we summarize its posterior distribution using the posterior mean together with the corresponding 95\% credible interval.

\begin{figure}[htbp]
\centering
    {\includegraphics[width=\textwidth]{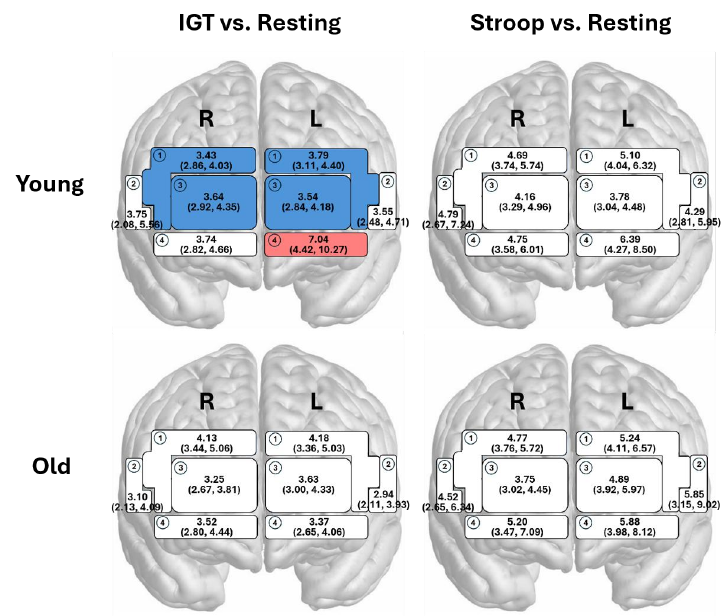}} 
    {\caption{ROI-specific importance maps for the IGT versus resting and Stroop versus resting comparisons in the younger and older cohorts. The values represent the posterior means of $\beta_{r_\ell}$, with the corresponding 95\% credible intervals shown in parentheses. Red ROIs indicate significantly larger $\beta_{r_\ell}$ values than the corresponding blue ROIs based on the 95\% credible intervals of their pairwise differences.}
    \label{fig:impor}}
\end{figure}

Figure~\ref{fig:impor} shows distinct patterns in the ROI-specific interaction parameters between the younger and older cohorts. For the older cohort, no significant differences among ROIs are observed for either the IGT or Stroop comparison, with the 95\% credible intervals overlapping for every pair
of ROIs. This lack of regional differentiation can be interpreted in the context of age-related neural dedifferentiation \citep{park2004aging}. Neural dedifferentiation refers to a reduction in the functional specificity or selectivity of neural responses with aging, such that neural representations become less differentiated across ROIs or conditions. MVPA studies comparing patterns of brain activity between younger and older adults have reported similar age-related reductions in neural distinctiveness \citep{carp2011age}. Moreover, MVPA-based measures of memory specificity have provided evidence of neural dedifferentiation in older adults in both the PFC and hippocampus \citep{abdulrahman2017dopamine}, suggesting that age-related dedifferentiation may extend across brain regions involved in higher-order cognitive and memory processes. 

Although the older cohort shows no significant ROI-specific differences, the estimated interaction parameters exhibit different bilateral patterns across tasks. In the IGT comparison, the left and right ROIs show relatively similar values of $\beta_{r_\ell}$, whereas greater left--right differences are observed in the Stroop comparison. The relatively similar bilateral interaction strengths, particularly for the IGT comparison, are broadly consistent with the hemispheric asymmetry reduction in older adults (HAROLD) model \citep{cabeza2002hemispheric}. The HAROLD model describes an age-related reduction in the lateralization of prefrontal activity during cognitive performance, with older adults tending to exhibit more bilateral patterns of prefrontal recruitment than younger adults. Consistent with this interpretation, studies using reward-based decision-making tasks similar to the IGT have reported greater bilateral and contralateral recruitment in older than in younger adults \citep{mccarrey2012age}, suggesting reduced hemispheric specialization during decision-making with aging. Previous studies have examined neural dedifferentiation and HAROLD using conventional measures of neural selectivity or hemispheric activation asymmetry. In contrast, our BDWI model quantifies regional importance through ROI-specific interaction parameters, enabling direct comparisons across ROIs.


The younger cohort exhibits significant regional differences in the IGT comparison. The interaction parameters for the DLPFC (ROI 1) and FPC (ROI 3) are significantly lower than that for the L-OFC (ROI 4L). This pattern may reflect functional differences among the OFC, DLPFC, and FPC regions. The OFC is closely involved in value representation, reward evaluation, and decision-making under uncertainty, processes that are particularly relevant to the IGT \citep{Bechara1994, bechara2000emotion}. In particular, \citet{lawrence2009distinct} reported greater activation of the lateral OFC during risky relative to safe deck choices in healthy adults performing the IGT. Accordingly, the stronger interaction associated with the L-OFC (ROI 4L) may indicate that signals from this ROI discriminate more strongly between IGT and resting conditions.
In contrast, the DLPFC is involved in working-memory-related processes \citep{li2010iowa}, whereas the FPC has been associated with higher-order integrative and control processes across multiple information sources \citep{ ramnani2004anterior}. These functional characteristics may be consistent with the significantly lower interaction parameters for ROI 1 and ROI 3, suggesting weaker condition-specific discrimination in these regions than in ROI 4L.

No significant pairwise differences among ROIs are observed in the younger cohort for the Stroop comparison. This finding may reflect differences in the neural demands of the two tasks. Whereas the IGT strongly engages value-evaluation processes for which the OFC has a relatively specialized role, Stroop performance relies on a broader cognitive-control network involving multiple prefrontal regions, including the DLPFC and VLPFC. This interpretation is consistent with the superordinate cognitive control network described by \citet{niendam2012meta}, in which diverse executive-control tasks recruit a distributed set of frontal and parietal regions.

\subsection{Connectivity Analysis}

We further investigate the pairwise connectivity structure based on the distance-based weights in equation~\eqref{eq:connectivity} by presenting the posterior mean and $95\%$ credible interval of $\widetilde C_{jk}(\sigma)$ for each pair of ROIs $j$ and $k$. Figure~\ref{fig:connec} illustrates the top five ROI pairs for each dataset, highlighting those with the strongest pairwise connectivity. The complete connectivity graph across all ROI pairs is provided in Section~3 of the Supplementary Material. 
\begin{figure}[htbp]%
\centering
{\includegraphics[width=\textwidth]{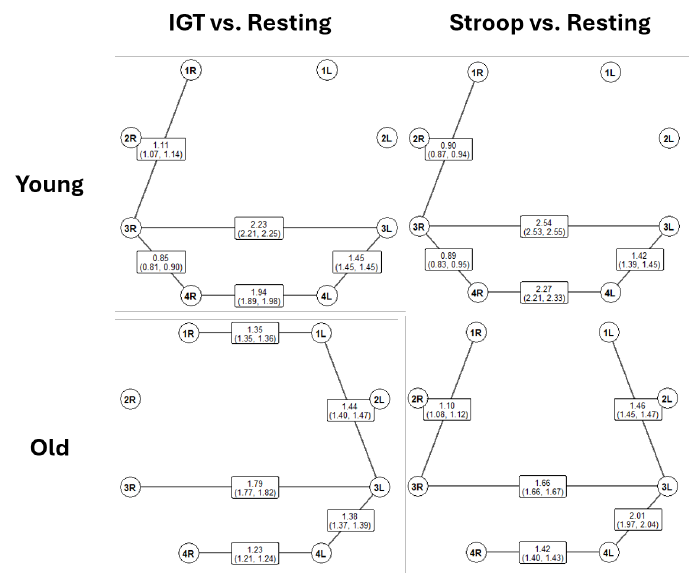}} 
{\caption{The five ROI pairs with the highest posterior mean distance-weighted connectivity for the IGT versus resting and Stroop versus resting comparisons in the younger and older cohorts. Each annotation gives the posterior mean $\widetilde C_{jk}(\sigma)$ and $95\%$ credible interval, defined in equation~(\ref{eq:connectivity}), for the corresponding ROI pairs.}
\label{fig:connec}}
\end{figure}

Several estimated model-based connectivity patterns can be interpreted in terms of the known functional and hemispheric organization of the prefrontal cortex. First, the bilateral connectivity between the left and right FPC (ROI 3) and OFC (ROI 4) is generally stronger than that between the corresponding DLPFC (ROI 1) and VLPFC (ROI 2). This pattern is broadly consistent with previously reported regional differences in interhemispheric organization of the prefrontal cortex. Electrophysiological evidence from intracranial cortico-cortical evoked potential (CCEP) recordings has demonstrated strong homotopic connections involving the OFC, characterized by short interhemispheric response latencies \citep{starkweather2026rapid}. Complementary evidence from meta-analytic task-based co-activation data has also shown substantial regional variation in homotopic connectivity within the prefrontal cortex. In particular, regions near the medial OFC exhibit relatively strong homotopic connectivity, whereas lateral prefrontal regions, including the DLPFC and VLPFC, show comparatively weak homotopic connectivity \citep{mancuso2019homotopic}. While these studies characterize interhemispheric organization through electrophysiological responses or co-activation patterns, our model quantifies regional differences through model-based connectivity between functional observations.

Second, strong connectivity between the left FPC (ROI 3L) and left OFC (ROI 4L) is consistently observed across all settings. This pattern may be explained by their anatomical proximity and direct structural connections. In particular, the fronto-orbitopolar tract provides an associative pathway between the posterior OFC and anterior orbitofrontal and frontal polar regions, including connections between areas corresponding to the OFC and FPC \citep{catani2012short}. Functionally, the strong FPC--OFC connectivity may also be consistent with their complementary involvement in value-based decision making, with the OFC closely associated with value and outcome evaluation \citep{Bechara1994, bechara2000emotion} and the FPC with the integration of information across multiple cognitive processes \citep{ramnani2004anterior}. The consistently strong connectivity between ROI 3L and ROI 4L may therefore be consistent with both anatomical and functional coupling between these neighboring prefrontal regions. 

Finally, the older cohort generally exhibits stronger connectivity between corresponding left and right ROIs than the younger cohort. This age-related pattern is consistent with the reduced hemispheric lateralization described by the HAROLD model \citep{cabeza2002hemispheric}. As discussed above, older adults tend to show more bilateral patterns of prefrontal recruitment during cognitive processing, reflecting reduced hemispheric asymmetry with aging. The HAROLD model primarily concerns hemispheric activation asymmetry, whereas the present analysis characterizes bilateral organization through model-based connectivity.


An additional insight emerges when regional importance and model-based connectivity are considered together. Across all cases, the left FPC and left OFC exhibit strong distance-weighted connectivity. In contrast, the left OFC shows greater ROI-specific interaction strength than the FPC in the younger group. This finding suggests that strong model-based functional connectivity between brain regions does not necessarily imply comparable contributions to task-related differentiation. 

\section{Discussion}
\label{sec:discussion}

In this study, we proposed a unified probabilistic model for neuroimaging data that jointly quantifies ROI-specific importance and distance-weighted functional connectivity. To address the computational challenges posed by high-dimensional functional data and an intractable normalizing constant, we utilized basis expansion with auxiliary variable MCMC. Our proposed model demonstrated improved model fit compared with the baseline models across all task comparisons, highlighting the benefit of allowing regional heterogeneity. The importance analysis revealed significant differences among ROIs for the IGT comparison in the younger cohort and no significant differences in the older cohort. The connectivity analysis identified difference between age cohorts, including greater bilateral connectivity in the older cohort.

Our findings are consistent with age-related neural dedifferentiation and reduced hemispheric lateralization, suggesting that aging may be associated with less regionally differentiated and more bilaterally similar patterns of prefrontal activity. Furthermore, significant differences in regional importance between the left FPC and the left OFC emerged in the younger comparison. Beyond these neuroscientific implications, our model enables joint inference on regional contributions and inter-regional functional relationships while quantifying uncertainty in their estimation.

Although our study focused on fNIRS data, the proposed framework could be extended to other neuroimaging modalities, such as fMRI, with suitable representations of functional observations and spatial regions. Future research could address the computational cost of the auxiliary variable simulation
required at each MCMC iteration, particularly for large datasets. One direction is to investigate whether divide-and-conquer methods \citep{srivastava2018scalable, scott2016bayes} can be adapted to improve the scalability of posterior inference for our proposed model.

\section*{Acknowledgements}
The authors are grateful to anonymous reviewers for their careful reading and valuable comments. This work was supported by the National Research Foundation of Korea (NRF-2025S1A5C3A0200632411). Nial Friel's research was conducted with the financial support of Taighde Éireann – Research Ireland, under Grant number [23/RC/13506] at Rinn Artificial Intelligence.

\clearpage
\appendix

\section{Details of Experimental Tasks and Data Preprocessing}
\label{appendix_a}
\subsection{Participants}
Participants were recruited separately for the younger and older adult groups. The younger adult sample consisted of 43 participants recruited from four-year universities across South Korea. Eligible participants were current undergraduate students or college graduates between 19 and 33 years of age. The final younger adult sample ranged in age from 19 to 32 years (mean age = 23.37 years) and included 28 female participants.
For the older adult group, 42 participants were recruited through a senior program in Seongnam-si, Gyeonggi-do. Cognitive eligibility was assessed using the Korean version of the Mini-Mental State Examination (K-MMSE) \citep{kang1997validity}, with scores below 24 serving as an exclusion criterion. Of the 62 initially recruited participants, one was excluded due to suspected mild cognitive impairment (MCI), and one was excluded for not meeting the minimum age requirement of 65 years. The final older adult sample therefore consisted of 60 participants, ranging in age from 66 to 88 years (mean age = 74.58 years), of whom 53 were female.
Across both age groups, participants were excluded if they (a) had a current or past Axis I psychiatric disorder, including substance abuse or dependence; (b) were taking medication on an ongoing basis that could plausibly alter neural activity relative to the general population; or (c) had uncorrected visual impairment that could interfere with viewing the on-screen stimuli. All eligible participants provided written informed consent prior to participation.

\subsection{Experimental Design and Procedures}

For the older cohort, eligibility was first assessed using the Korean version of the Mini-Mental State Examination (K-MMSE) \citep{kang1997validity}, with participants scoring below 24 excluded from participation. Eligible participants provided written informed consent and subsequently completed a set of paper-based questionnaires.
Participants then completed a keyboard-based version of the Iowa Gambling Task (IGT). Task instructions were presented via slides, and participants responded using the index finger of one hand. The functional near-infrared spectroscopy (fNIRS) device was fitted according to a standardized manual. Before the main task, a 5-minute resting-state baseline was recorded while participants sat quietly with their eyes closed. Participants were informed when the resting period had concluded. They subsequently completed a 20-trial IGT practice block followed by the 100-trial main IGT block.
The Stroop task was administered after the IGT. Instructions were presented via slides, and participants responded using the keyboard arrow keys as quickly and accurately as possible. Participants completed a practice block until they were familiar with the response procedure, after which the main Stroop task was administered. The main task consisted of a 30-second congruent condition followed by a 30-second incongruent condition.

For the younger cohort, Eligible participants provided written informed consent and subsequently completed a computerized questionnaire battery using a mouse. The experimenter explained the initial instruction screen and informed participants that instructions would be presented before each subsequent scale. Participants then proceeded through the questionnaires independently and notified the experimenter upon completion.
Participants next completed a keyboard-based version of the IGT. Task instructions were presented via slides, and participants were instructed to respond using the index finger of one hand. After completing a 10-trial practice block, the fNIRS device was fitted according to a standardized manual, and the corresponding PsychoPy experimental file was configured. A 5-minute resting-state baseline was then recorded, during which participants were instructed to remain still and either maintain fixation or keep eyes closed without falling asleep. Following the baseline recording, participants completed the 200-trial main IGT block.
The Stroop task was subsequently administered using PsychoPy for stimulus presentation and response collection. Participants were instructed to respond using the keyboard as quickly as possible while maintaining accurate performance. The experimenter monitored participants during the practice block to ensure familiarity with the response procedure. The main Stroop task then followed the same structure as that used for the older cohort, consisting of a 30-second congruent condition followed by a 30-second incongruent condition.

\subsection{Tasks}
Participants completed a computerized version of the IGT using a keyboard, responding with the index finger of one hand. The reward and loss contingencies followed the original payoff scheme introduced by \cite{Bechara1994}. However, unlike the original paradigm, we did not impose a maximum number of selections from each deck. If the number of selections from a given deck exceeded the length of its predefined payoff sequence, the sequence was repeated from the beginning. 
Decks A and B were disadvantageous, providing a constant reward of 100 USD per selection but imposing relatively large losses, resulting in a net expected loss of 250 USD per 10 selections. Conversely, In contrast, Decks C and D were advantageous, providing a smaller constant reward of 50 USD per selection but imposing smaller cumulative losses, resulting in a net expected gain of 250 USD per 10 selections. The decks also differed in loss frequency: losses occurred on 5 of every 10 selections for Decks A and C and on 1 of every 10 selections for Decks B and D. To facilitate interpretation of the monetary outcomes by Korean participants, all monetary values were proportionally rescaled to Korean Won (KRW). For example, the 100 USD reward associated with Decks A and B was represented as 10,000 KRW.
The behavioral data were collected as part of a concurrent fNIRS study. Accordingly, we used a fixed trial duration to maintain a consistent temporal structure for the neuroimaging experiment. Previous research has shown that imposing time constraints on the IGT does not substantially alter behavioral patterns or decision-making trajectories \citep{bowman2005artificial}. Because younger adults responded substantially faster than older adults, the trial duration was adjusted by age group to avoid prolonged inactive periods within trials while maintaining a comparably constrained task structure. Older adults completed 100 trials with a fixed duration of 6 seconds per trial, whereas younger adults completed 200 trials with a fixed duration of 4 seconds per trial.
If a participant failed to make a selection within the response window, a warning message instructing the participant to respond within the allotted time was displayed for the remainder of the trial. The trial was recorded as a non-response (NA) and excluded from the behavioral analysis. No outcome feedback was presented for non-response trials; consequently, these trials provided no new information about the reward or loss associated with any deck. In the subsequent behavioral analysis, deck-specific value representations were therefore carried forward unchanged until the next valid response.

Selective attention and cognitive interference were assessed using a computerized Stroop task \citep{Stroop1935}, which was administered using the same procedure in both age groups. On each trial, a color word was presented in either a congruent condition (e.g., the word 'RED' displayed in red) or an incongruent condition (e.g., the word 'RED' displayed in blue). Participants were instructed to identify the color in which the word was displayed as quickly and accurately as possible by pressing the corresponding directional arrow key on a keyboard. Each of the four arrow keys was mapped to one of four response colors: red, blue, green, and yellow.
Before the main task, participants completed a fixed 16-trial practice block under the supervision of the experimenter. The main task consisted of a 30-second congruent block followed by a 30-second incongruent block, resulting in a total task duration of 60 seconds.

Response time (RT; ms), accuracy (proportion correct), and the number of completed trials were recorded separately for each block. To quantify the Stroop interference effect, an interference score was calculated for each measure as the difference between performance in the incongruent and congruent blocks (incongruent $-$ congruent; \citealp{cothran2008comparison, dittrich2012selective}). This yielded three measures: accuracy interference, trial-count interference, and RT interference. A positive RT interference score indicates slower responses in the incongruent condition, whereas negative accuracy and trial-count interference scores indicate lower accuracy and fewer completed trials, respectively, in the incongruent condition. Thus, larger positive values for RT interference and larger negative values for accuracy and trial-count interference reflect greater Stroop interference.

\subsection{fNIRS Data Acquisition}

Cerebral hemodynamic activity was recorded using a continuous-wave functional near-infrared spectroscopy (fNIRS) system (NIRSIT; OBELAB Inc., Seoul, Republic of Korea). The portable, wireless system uses two wavelengths of near-infrared light (approximately 780 and 850 nm) to measure changes in oxygenated and deoxygenated hemoglobin concentrations (HbO and HbR) in cortical tissue. The system comprises 24 laser diodes and 32 photodiodes, providing 48 measurement channels with a source--detector separation of 3 cm and a sampling frequency of 8.138 Hz. Measurements covered the bilateral dorsolateral prefrontal cortex (DLPFC), frontopolar cortex (FPC), ventrolateral prefrontal cortex (VLPFC), and orbitofrontal cortex (OFC). Data acquisition was performed using NIRSIT SCAN (OBELAB Inc., Seoul, Republic of Korea), the manufacturer's dedicated acquisition software.
fNIRS measurements were conducted in a quiet, controlled environment with minimal interference from strong artificial light sources or electromagnetic signals. Participants were seated comfortably and instructed to remain still during data acquisition. Before positioning the NIRSIT probes over the forehead, care was taken to ensure that hair did not obstruct contact at the forehead and temporal regions. The system's built-in gain calibration procedure was then performed automatically to optimize signal quality before recording.

The fNIRS data were preprocessed using a standardized sequence of quality-control and signal-processing procedures. First, invalid samples were identified at the raw intensity level. When five or fewer invalid samples occurred consecutively, they were replaced using nearest-neighbor interpolation; channels containing more than five consecutive invalid samples were rejected. Additional channel-quality criteria were then applied to the raw intensity data. Channels were rejected if their median intensity was below 30 \citep{Yucel2021}, if their coefficient of variation exceeded 20\% \citep{pfeifer2018signal}, or if identical consecutive values occurred over more than 5\% of the entire time series, indicating potential signal saturation. An additional quality-control criterion was applied after conversion to hemoglobin concentrations: channels exhibiting an extreme negative correlation between the HbO and HbR time series ($r \leq -0.9$), indicative of a potential mirroring artifact, were rejected \citep{takizawa2014neuroimaging}.

Following the initial quality-control procedures, the retained raw intensity data were converted to optical density \citep{Yucel2021}, and motion artifacts were corrected using Temporal Derivative Distribution Repair (TDDR) \citep{fishburn2019temporal}. The corrected optical density data were subsequently converted to changes in hemoglobin concentration using the modified Beer--Lambert law \citep{delpy1988estimation, cope1988system}. Following the ISO guideline, concentration changes were calculated without differential pathlength correction and were therefore expressed in units of $\mathrm{mm}\cdot\mathrm{mM}$. Molar extinction coefficients based on the data reported by Moaveni were used for this conversion \citep{zhao2017optimal}. The resulting hemoglobin concentration signals were then digitally filtered using a fourth-order Butterworth bandpass filter with cutoff frequencies of 0.01 and 0.2 Hz to attenuate low-frequency drift and higher-frequency physiological and measurement noise \citep{Yucel2021}. Finally, channels rejected during the quality-control procedures were reconstructed using a predefined padding procedure. For each rejected channel, replacement values were first obtained from designated backup channels with the same source--detector distance and corresponding target location. When no suitable backup channel was available, the rejected channel was replaced using the mean signal of the channels belonging to the same Brodmann area group.

\section{Simulation Study}
\label{appendix_b}
In this section, we conducted a simulation study to assess whether the proposed BDWI model can accurately recover the model parameters and provide appropriate uncertainty quantification. To retain the structure of the functional neuroimaging data used in the application, we based the simulation on the fNIRS observations from the older cohort for the IGT versus resting state comparison. The original fNIRS trajectories were represented using the same basis expansion as in the real-data analysis, and the resulting basis coefficients were used as the covariates $\mathbf X$. A fixed subset of observations ($n=769$) was used throughout the simulation study, together with their corresponding ROI memberships. Thus, $\mathbf X$ and the ROI indicators were held fixed across simulation replications. Given $\mathbf X$, the ROI memberships, and true parameter values $(\beta_1^{\mathrm{true}},\ldots,\beta_8^{\mathrm{true}})',$ and $\sigma^{\mathrm{true}}$, task labels were generated from the BDWI model using Gibbs sampling. For each replication, the Gibbs sampler was run for 1,000 sweeps before retaining the simulated task-label configuration. This procedure was repeated independently for 100 simulation replications.

For each simulated dataset, the BDWI model was fitted using the same Bayesian inference procedure as in the real-data analysis. Posterior inference was carried out using the double Metropolis--Hastings algorithm with auxiliary Gibbs updates. For each parameter, we recorded the posterior mean and the $95\%$ posterior credible interval.
The simulation results are summarized in Table~\ref{tab:simulation_bdwi_loop10} and ~\ref{tab:simulation_bdwi_loop20}. Table~\ref{tab:simulation_bdwi_loop10} presents the results obtained using 10 Gibbs sweeps for the BDWI procedure, while Table~\ref{tab:simulation_bdwi_loop20} presents the corresponding results using 20 Gibbs sweeps. We observe accurate parameter recovery for both $L=10$ and $L=20$, suggesting that $L=10$ sweeps are sufficient for accurate inference.

\begin{table}[!ht]
\centering
\caption{Simulation results for the BDWI model based on 100 replications, using 10 Gibbs sweeps. Posterior mean denotes the average of the posterior mean estimates across replications, and coverage denotes the empirical coverage probability of the 95\% posterior credible intervals.}
\label{tab:simulation_bdwi_loop10}
\begin{tabular}{lcccc}
\hline
Parameter & True & Posterior mean & RMSE & Coverage  \\
\hline
$\beta_1$ & 4.30 & 3.99 & 0.89 & 0.91\\
$\beta_2$ & 3.00 & 2.98 & 0.84 & 0.95\\
$\beta_3$ & 3.60 & 3.58 & 0.73 & 0.96\\
$\beta_4$ & 3.40 & 3.47 & 1.00 & 0.94\\
$\beta_5$ & 4.10 & 3.89 & 0.85 & 0.92\\
$\beta_6$ & 3.10 & 3.40 & 0.92 & 0.96\\
$\beta_7$ & 3.30 & 3.26 & 0.66 & 0.98\\
$\beta_8$ & 3.60 & 3.73 & 0.94 & 0.95\\
$\sigma$  & 0.20 & 0.21 & 0.01 & 0.99\\
\hline
\end{tabular}
\end{table}

\begin{table}[!ht]
\centering
\caption{Simulation results for the BDWI model based on 100 replications, using 20 Gibbs sweeps. Posterior mean denotes the average of the posterior mean estimates across replications, and coverage denotes the empirical coverage probability of the 95\% posterior credible intervals.}
\label{tab:simulation_bdwi_loop20}
\begin{tabular}{lcccc}
\hline
Parameter & True & Posterior mean & RMSE & Coverage  \\
\hline
$\beta_1$ & 4.30 & 3.94 & 0.88 & 0.92\\
$\beta_2$ & 3.00 & 2.97 & 0.84 & 0.97\\
$\beta_3$ & 3.60 & 3.55 & 0.69 & 0.96\\
$\beta_4$ & 3.40 & 3.45 & 0.98 & 0.93\\
$\beta_5$ & 4.10 & 3.85 & 0.83 & 0.91\\
$\beta_6$ & 3.10 & 3.39 & 0.91 & 0.96\\
$\beta_7$ & 3.30 & 3.22 & 0.65 & 0.98\\
$\beta_8$ & 3.60 & 3.71 & 0.93 & 0.95\\
$\sigma$  & 0.20 & 0.20 & 0.01 & 0.99\\
\hline
\end{tabular}
\end{table}

\section{Extra Results}
\label{appendix_c}

\subsection{Connectivity Map}

\begin{figure}[htbp]
    \centering
    \includegraphics[width=\textwidth]{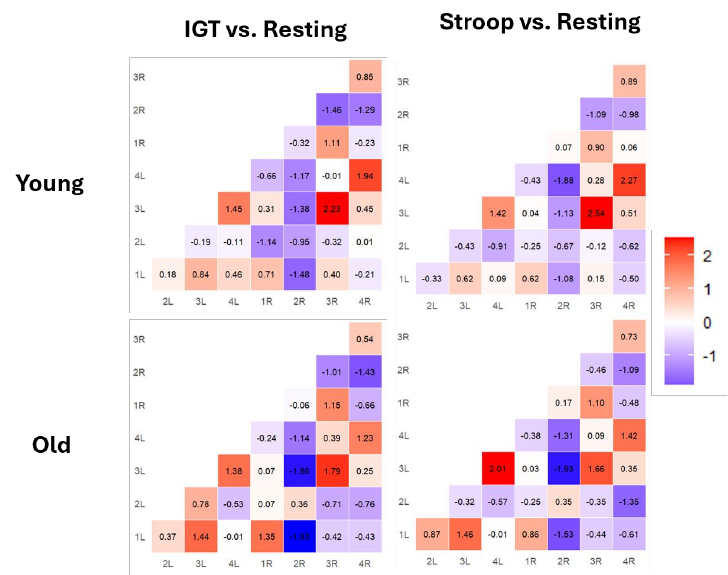} 
    \caption{ROI-specific connectivity graph for the IGT versus resting and Stroop versus resting comparisons in the younger and older cohorts. Here each value and interval correspond to the posterior mean $\widetilde C_{jk}(\sigma)$ and $95\%$ credible interval, defined in equation~(5), for a given pair of locations $j$ and $k$.}
    \label{fig:connec}
\end{figure}

\bibliography{ref}

\end{document}